\documentclass[twocolumn]{aastex63}

\received{}
\revised{}
\accepted{}
\submitjournal{ApJL}

\shorttitle{ASKAP observations of S190814bv}
\shortauthors{Dobie et al.}
\graphicspath{{./}{figures/}}

\makeatletter
\newcommand\footnoteref[1]{\protected@xdef\@thefnmark{\ref{#1}}\@footnotemark}
\makeatother

\begin{document}

\title{An ASKAP search for a radio counterpart to the first high-significance neutron star-black hole merger LIGO/Virgo S190814bv}

\correspondingauthor{Dougal Dobie}
\email{ddob1600@uni.sydney.edu.au, tara.murphy@sydney.edu.au}

\author[0000-0003-0699-7019]{Dougal Dobie}
\affiliation{Sydney Institute for Astronomy, School of Physics, University of Sydney, Sydney, New South Wales 2006, Australia.}
\affiliation{ATNF, CSIRO Astronomy and Space Science, PO Box 76, Epping, New South Wales 1710, Australia}
\affiliation{ARC Centre of Excellence for Gravitational Wave Discovery (OzGrav), Hawthorn, Victoria, Australia}

\author[0000-0001-8026-5903]{Adam Stewart}
\affiliation{Sydney Institute for Astronomy, School of Physics, University of Sydney, Sydney, New South Wales 2006, Australia.}

\author[0000-0002-2686-438X]{Tara Murphy}
\affiliation{Sydney Institute for Astronomy, School of Physics, University of Sydney, Sydney, New South Wales 2006, Australia.}
\affiliation{ARC Centre of Excellence for Gravitational Wave Discovery (OzGrav), Hawthorn, Victoria, Australia}

\author[0000-0002-9994-1593]{Emil Lenc}
\affiliation{ATNF, CSIRO Astronomy and Space Science, PO Box 76, Epping, New South Wales 1710, Australia}

\author[0000-0002-2066-9823]{Ziteng Wang}
\affiliation{Sydney Institute for Astronomy, School of Physics, University of Sydney, Sydney, New South Wales 2006, Australia.}
\affiliation{ATNF, CSIRO Astronomy and Space Science, PO Box 76, Epping, New South Wales 1710, Australia}

\author[0000-0001-6295-2881]{David~L.~Kaplan}
\affiliation{Center for Gravitation, Cosmology, and Astrophysics, Department of Physics, University of Wisconsin-Milwaukee, P.O. Box 413, Milwaukee, WI 53201, USA}

\author[0000-0002-8977-1498]{Igor~Andreoni}
\affiliation{Division of Physics, Mathematics, and Astronomy, California Institute of Technology, Pasadena, CA 91125, USA}

\author[0000-0003-4417-5374]{Julie Banfield}
\affiliation{ATNF, CSIRO Astronomy and Space Science, PO Box 76, Epping, New South Wales 1710, Australia}

\author[0000-0003-1095-8194]{Ian Brown}
\affiliation{Center for Gravitation, Cosmology, and Astrophysics, Department of Physics, University of Wisconsin-Milwaukee, P.O. Box 413, Milwaukee, WI 53201, USA}

\author[0000-0001-8104-3536]{Alessandra Corsi}
\affiliation{Department of Physics and Astronomy, Texas Tech University, Box 1051, Lubbock, TX 79409, USA}

\author[0000-0002-8989-0542]{Kishalay De}
\affiliation{Division of Physics, Mathematics, and Astronomy, California Institute of Technology, Pasadena, CA 91125, USA}

\author[0000-0003-3461-8661]{Daniel~A.~Goldstein}
\altaffiliation{Hubble Fellow}
\affiliation{Division of Physics, Mathematics, and Astronomy, California Institute of Technology, Pasadena, CA 91125, USA}

\author[0000-0002-7083-4049]{Gregg Hallinan}
\affiliation{Cahill Center for Astronomy \& Astrophysics, Caltech, Pasadena CA, USA}

\author{Aidan Hotan}
\affiliation{ATNF, CSIRO Astronomy and Space Science, PO Box 1130, Bentley, WA 6102, Australia}

\author[0000-0002-2502-3730]{Kenta Hotokezaka}
\affiliation{Department of Astrophysical Sciences, Princeton University, 4 Ivy Lane, Princeton, NJ 08544, USA}
\affiliation{Research Center for the Early Universe, Graduate School of Science, University of Tokyo, Bunkyo-ku, Tokyo 113-0033, Japan}

\author[0000-0002-3850-6651]{Amruta D.\ Jaodand}
\affiliation{Division of Physics, Mathematics, and Astronomy, California Institute of Technology, Pasadena, CA 91125, USA}

\author{Viraj Karambelkar}
\affiliation{Division of Physics, Mathematics, and Astronomy, California Institute of Technology, Pasadena, CA 91125, USA}

\author[0000-0002-5619-4938]{Mansi M. Kasliwal}
\affiliation{Division of Physics, Mathematics, and Astronomy, California Institute of Technology, Pasadena, CA 91125, USA}

\author{David McConnell}
\affiliation{ATNF, CSIRO Astronomy and Space Science, PO Box 76, Epping, New South Wales 1710, Australia}

\author[0000-0002-2557-5180]{Kunal Mooley}
\affiliation{Cahill Center for Astronomy \& Astrophysics, Caltech, Pasadena CA, USA}

\author[0000-0002-3005-9738]{Vanessa A. Moss}
\affiliation{ATNF, CSIRO Astronomy and Space Science, PO Box 76, Epping, New South Wales 1710, Australia}
\affiliation{Sydney Institute for Astronomy, School of Physics, University of Sydney, Sydney, New South Wales 2006, Australia.}

\author[0000-0001-8684-2222]{Jeffrey A. Newman}
\affiliation{Department of Physics and Astronomy, University of Pittsburgh, 3941 O’Hara Street, Pittsburgh, PA 15260, USA}
\affiliation{Pittsburgh Particle Physics, Astrophysics, and Cosmology Center (PITT PACC), Pittsburgh, PA 15260, USA}

\author[0000-0001-8472-1996]{Daniel A. Perley}
\affil{Astrophysics Research Institute, Liverpool John Moores University,\\ IC2, Liverpool Science Park, 146 Brownlow Hill, Liverpool L3 5RF, UK}

\author[0000-0003-4451-4444]{Abhishek Prakash}
\affil{IPAC, California Institute of Technology
1200 E California Boulevard, Pasadena, CA 91125, USA}

\author[0000-0003-1575-5249]{Joshua Pritchard}
\affiliation{Sydney Institute for Astronomy, School of Physics, University of Sydney, Sydney, New South Wales 2006, Australia.}

\author[0000-0002-1136-2555]{Elaine M. Sadler}
\affiliation{Sydney Institute for Astronomy, School of Physics, University of Sydney, Sydney, New South Wales 2006, Australia.}
\affiliation{ATNF, CSIRO Astronomy and Space Science, PO Box 76, Epping, New South Wales 1710, Australia}

\author{Yashvi Sharma}
\affiliation{Division of Physics, Mathematics, and Astronomy, California Institute of Technology, Pasadena, CA 91125, USA}

\author{Charlotte Ward}
\affiliation{Department of Astronomy, University of Maryland, College Park, MD 20742, USA}

\author[0000-0003-1160-2077]{Matthew Whiting}
\affiliation{ATNF, CSIRO Astronomy and Space Science, PO Box 76, Epping, New South Wales 1710, Australia}

\author[0000-0001-5381-4372]{Rongpu Zhou}
\affiliation{Lawrence Berkeley National Laboratory, 1 Cyclotron Road, Berkeley, CA 94720, USA}

\begin{abstract}
We present results from a search for a radio transient associated with the LIGO/Virgo source S190814bv, a likely neutron star-black hole (NSBH) merger, with the Australian Square Kilometre Array Pathfinder. We imaged a $30\,{\rm deg}^2$ field at $\Delta T=2$, 9 and 33 days post-merger at a frequency of 944\,MHz, comparing them to reference images from the Rapid ASKAP Continuum Survey observed 110 days prior to the event. Each epoch of our observations covers $89\%$ of the LIGO/Virgo localisation region. We conducted an untargeted search for radio transients in this field, resulting in 21 candidates. For one of these, \object[AT2019osy]{AT2019osy}, we performed multi-wavelength follow-up and ultimately ruled out the association with S190814bv. All other candidates are likely unrelated variables, but we cannot conclusively rule them out. We discuss our results in the context of model predictions for radio emission from neutron star-black hole mergers and place constrains on the circum-merger density and inclination angle of the merger. This survey is simultaneously the first large-scale radio follow-up of an NSBH merger, and the most sensitive widefield radio transients search to-date.
\end{abstract}


\section{Introduction}
On 14 August 2019 the LIGO and Virgo collaborations detected the compact binary merger S190814bv\footnote{\url{https://gracedb.ligo.org/superevents/S190814bv/view/}} with the LIGO Livingston (L1), LIGO Hanford (H1) and Virgo (V1) gravitational wave detectors \citep{GCN25324}. The event was classified as a neutron star--black hole (NSBH) merger, where the lighter component has a mass $<3\,M_\odot$, and the heavier component has a mass $>5\,M_\odot$, \citep{GCN25333}. The accuracy of this classification is dependent on the physical upper-limit for neutron star mass which is not well constrained, but may be less than the above definition \citep{2019MNRAS.488.5020Z,2019NatAs.tmp..439C}. The probability of there being matter outside the remnant object is $<1$\% \citep{GCN25324}, therefore the expected nature of any electromagnetic radiation from the merger (if any) is unclear.

The preferred skymap (\texttt{LALInference.v1.fits.gz}) has a 90\% localisation region of 23\,$\deg^2$ and a sky-averaged distance estimate of $267\pm52$\,Mpc. High-energy observations \citep{GCN25323,GCN25326,GCN25327,GCN25329,GCN25341} find no evidence for a coincident short gamma-ray burst (GRB). Optical observations found numerous candidate counterparts that have since been ruled out with further photometric and spectroscopic observations (Andreoni et al. in prep.). 

While the low probability of remnant matter \citep{GCN25333} may suggest that the merger produced no electromagnetic counterpart, the lack of optical counterparts may also be explained by intrinsic factors such as inclination angle, mass ratio, remnant lifetime or a lack of polar ejecta \citep{2017Natur.551...80K}, or extrinsic factors like dust-obscuration. In this case, radio emission may be the only way to localise this event.

We performed follow-up of S190814bv with the Australian Square Kilometre Array Pathfinder \citep[ASKAP;][]{2008ExA....22..151J}. In Section \ref{s_transients} we discuss our untargeted radio transients search. In Section \ref{s_follow} we summarise multi-wavelength follow-up of candidate counterpart \object[AT2019osy]{AT2019osy} that was initially detected in this search.   

\section{Observations \& Data Reduction}
\begin{figure*}
    \centering
    \includegraphics[width=\linewidth]{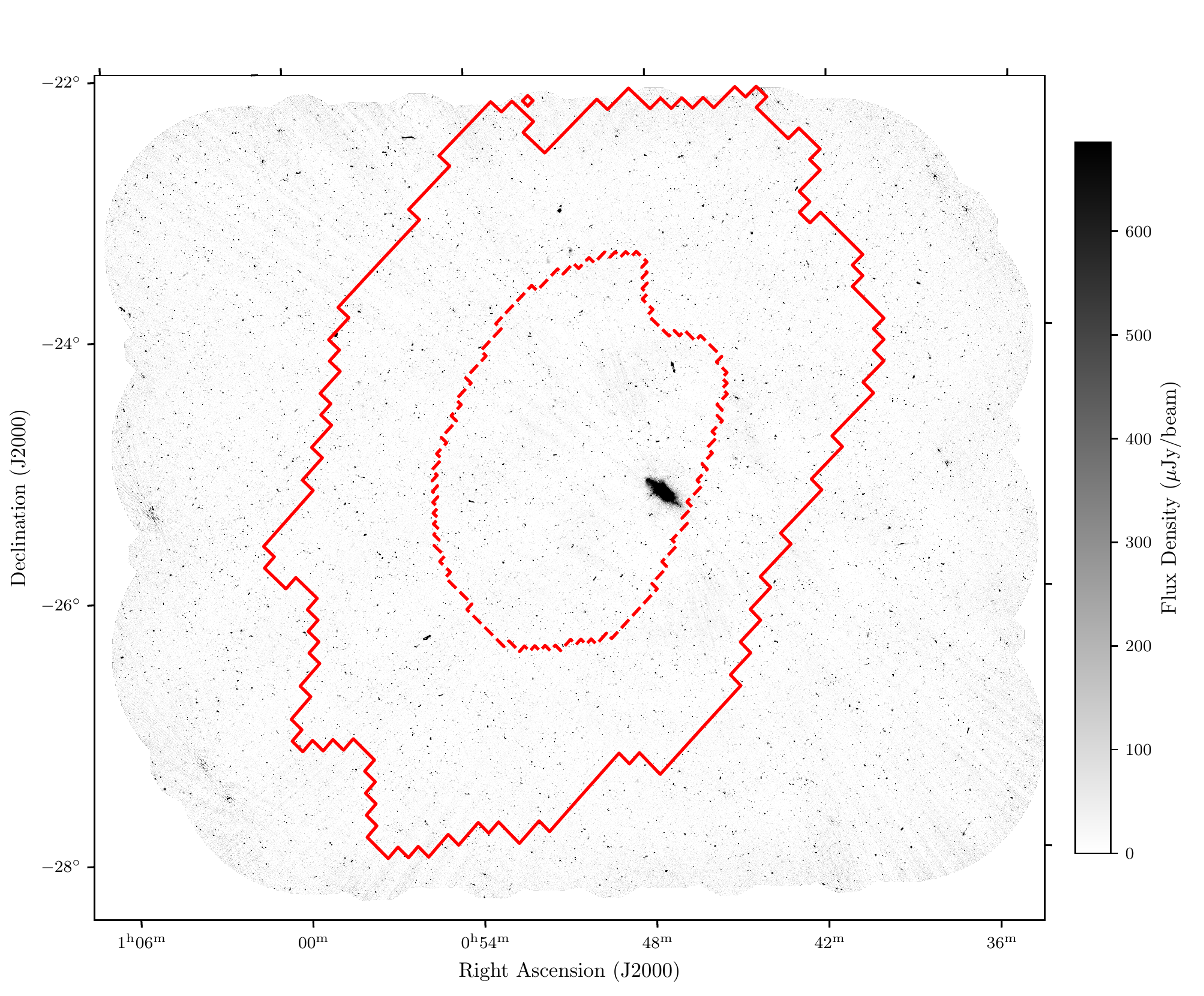}
    \caption{ASKAP image of the localisation region of S190814bv centered on 00:50:37.5, $-$25:16:57.371 observed 2 days post-merger. The 30$\deg^2$ field of view covers $\sim 89\%$ of the localisation region, with 50\% (90\%) contours shown in red dashed (solid) lines. The large object near the centre of the image is the radio-emitting starburst galaxy NGC\,253. Note: there is a secondary lobe of the localisation towards the south-east that is outside the ASKAP footprint.}
    \label{fig:askap_img}
\end{figure*}

We observed a target field centred on (J2000) coordinates $\alpha=00^{\rm h}50^{\rm m}37\fs5$, $\delta=-25\degr16\arcmin57\fs37$ at $\Delta T=2$, 9 and 33\,days post-merger with ASKAP. This target field, shown in Figure~\ref{fig:askap_img} at $\Delta T=2$\,days, covers $89\%$ of the skymap probability. 

Table~\ref{tab:obs_descrip} gives a summary of our ASKAP observations. Data were observed using 36 beams arranged in a closepack36 footprint\footnote{For more information on ASKAP beam-forming, see: \url{https://confluence.csiro.au/display/askapsst/}} with beam spacing of 0.9 degrees. The field was tracked for a nominal time of 10.5\,hrs and 288\,MHz of bandwidth was recorded with a center frequency of 944\,MHz. Typical sensitivity was $\sim 39\,\mu$Jy with a beam size of $\sim 12\arcsec$.

We imaged the data with the ASKAPsoft pipeline version 0.24.4 \citep{2017ASPC..512..431W}, using a set of parameters optimised for deep continuum fields. Each beam was imaged independently and then combined using a linear mosaic. Multi-frequency synthesis with two Taylor terms was used, along with Multi-scale CLEAN using scales up to 27 pixels in size. Visibilities were weighted using Wiener preconditioning with a robustness parameter of zero.
Two major cycles of self--calibration were used to refine the antenna gain solutions derived from observations of \object[PKS B1934-638]{PKS~B1934$-$638} in each beam \citep[see][for a description of the ASKAP beamforming and calibration process]{2016PASA...33...42M}. We also used pre-release data from the 888~MHz Rapid ASKAP Continuum Survey (RACS\footnote{\url{https://www.atnf.csiro.au/content/racs}}) as a reference epoch. 

\begin{deluxetable*}{clcccccc}
\tablecaption{Details of our ASKAP observations for each scheduling block ID (SBID). All observations were carried out with 288\,MHz of bandwidth centered on a frequency of 944\,MHz and 33 of 36 antennas. Typically 26\% of the data was flagged due to RFI or correlator drop-outs. The ASKAP images from our follow-up observations are available from the CSIRO ASKAP Science Data Archive\footnote{\url{https://casda.csiro.au/}} under project code AS111.\label{tab:obs_descrip}}
\tablehead{
\colhead{Epoch} &
\colhead{SBID} &
\colhead{Start} &
\colhead{Int. time} &
\colhead{$\Delta$T} &
\colhead{\% Flagged} &
\colhead{Sensitivity} &
\colhead{Beam Size}\\
\colhead{} &
\colhead{} &
\colhead{(UTC)} &
\colhead{(h:m:s)} &
\colhead{(d)} &
\colhead{} &
\colhead{($\mu$Jy)}\\
}
\startdata
0 & 8582 & 2019-04-27 04:59:14 & 00:15:00 & $-$110 & 26  & 270 & $10.2\arcsec\times14.9\arcsec$\\
1 & 9602 & 2019-08-16 14:10:27 & 10:39:25 & 2 & 25 & 35  & $10.0\arcsec\times12.3\arcsec$\\
2 & 9649 & 2019-08-23 13:42:59 & 10:39:01 & 9 & 26 & 39  & $11.8\arcsec\times12.4\arcsec$\\
3 & 9910 & 2019-09-16 12:08:34 & 10:38:42 & 33 & 32 & 39  & $9.8\arcsec\times12.1\arcsec$\\
\enddata
\end{deluxetable*}

The astrometric accuracy and flux scaling of each epoch is consistent with every other epoch. The median flux ratio of compact sources for any two of the ASKAP observations is consistent with 1 within uncertainties. The median RA offset is 0.09--0.36\arcsec and the median declination offset is 0.02--0.2\arcsec (smaller than the pixel size) with a typical standard deviation of 0.7\arcsec and 0.6\arcsec respectively.

\begin{deluxetable*}{lrlcccccccc}
\rotate
\tablecaption{Candidate counterparts from an untargeted search of the S190814bv localisation region. Non-detections are denoted by $3\sigma$ upper-limits based on the local noise measured by \texttt{BANE} \citep{2018PASA...35...11H}. The angular separation and redshift of the corresponding optical source are shown.\label{tab:candidates}}
\tablecolumns{9}
\tablehead{
\colhead{Name} &
\colhead{RA} &
\colhead{Dec} &
\colhead{$S_0$} &
\colhead{$S_1$} &
\colhead{$S_2$} &
\colhead{$S_3$} &
\colhead{$V_{\rm int}$} &
\colhead{$\eta_{\rm int}$} &
\colhead{offset} &
\colhead{z} \\
\colhead{} &
\colhead{($\deg$)} &
\colhead{($\deg$)} &
\colhead{(mJy)} &
\colhead{(mJy)} &
\colhead{(mJy)} &
\colhead{(mJy)} &
\colhead{} &
\colhead{} &
\colhead{($\arcsec$)} &
\colhead{}\\
}
\startdata
ASKAP~J004033.2$-$233530 & $10.13813$ & $-23.5917$ &  $4.700 \pm 0.454$ &  $4.517 \pm 0.062$ &  $4.732 \pm 0.069$ &  $6.648 \pm 0.068$ &  0.22 &     306 &  -- &                                -- \\
 ASKAP~J004054.8$-$273246 & $10.22816$ & $-27.5463$ &            $< 1.1$ &  $0.498 \pm 0.069$ &  $0.525 \pm 0.076$ &  $0.272 \pm 0.078$ &  0.32 &    3.29 & 13.4 &                   $0.19 \pm 0.05$ \\
 ASKAP~J004150.3$-$270632 & $10.45977$ & $-27.1090$ &            $< 1.0$ &  $0.656 \pm 0.058$ &  $0.536 \pm 0.063$ &  $0.436 \pm 0.064$ &  0.20 &    3.32 &  -- &                                -- \\
 ASKAP~J004424.5$-$265522 & $11.10216$ & $-26.9230$ &            $< 1.2$ &  $0.281 \pm 0.055$ &  $0.437 \pm 0.060$ &  $0.475 \pm 0.060$ &  0.26 &    3.26 &  -- &                                -- \\
 ASKAP~J004825.7$-$264137 & $12.10704$ & $-26.6937$ &           $< 0.75$ &  $0.384 \pm 0.053$ &  $0.615 \pm 0.057$ &  $0.614 \pm 0.057$ &  0.25 &    5.94 &  -- &                                -- \\
 ASKAP~J004916.8$-$270745 & $12.32005$ & $-27.1292$ &           $< 0.88$ &  $0.586 \pm 0.049$ &  $0.725 \pm 0.053$ &  $0.954 \pm 0.055$ &  0.25 &    12.6 & 16.8 &  $0.38 \pm 0.13$\tablenotemark{a} \\
 ASKAP~J005234.9$-$264144 & $13.14558$ & $-26.6956$ &           $< 0.73$ &  $0.379 \pm 0.050$ &  $0.380 \pm 0.055$ &  $0.226 \pm 0.054$ &  0.27 &    2.75 &  -- &                                -- \\
 ASKAP~J005304.8$-$255451 & $13.27001$ & $-25.9144$ &            $< 1.1$ &  $0.230 \pm 0.050$ &  $0.375 \pm 0.054$ &  $0.214 \pm 0.053$ &  0.33 &    2.75 &  -- &                                -- \\
 ASKAP~J005426.1$-$253833 & $13.60866$ & $-25.6425$ &           $< 0.72$ &  $0.274 \pm 0.053$ &  $0.487 \pm 0.059$ &  $0.273 \pm 0.059$ &  0.36 &    4.51 & 17.9 &                   $0.33 \pm 0.11$ \\
 ASKAP~J005434.6$-$280235 & $13.64412$ & $-28.0431$ &           $< 0.70$ &  $3.399 \pm 0.097$ &  $1.337 \pm 0.103$ &  $1.264 \pm 0.104$ &  0.61 &     149 & 11.5 &                   $0.21 \pm 0.11$ \\
 ASKAP~J005523.7$-$250403 & $13.84868$ & $-25.0675$ &           $< 0.86$ &  $0.972 \pm 0.053$ &  $0.753 \pm 0.060$ &  $0.669 \pm 0.060$ &  0.20 &    7.85 &  -- &                                -- \\
 ASKAP~J005547.4$-$270433 & $13.94764$ & $-27.0759$ &           $< 0.80$ &  $0.399 \pm 0.055$ &  $0.598 \pm 0.059$ &  $0.557 \pm 0.059$ &  0.20 &    3.45 &  0.1 &         $0.0733$\tablenotemark{b} \\
 ASKAP~J005606.9$-$255300 & $14.02875$ & $-25.8835$ &           $< 0.80$ &  $0.623 \pm 0.052$ &  $0.899 \pm 0.059$ &  $1.011 \pm 0.059$ &  0.24 &    13.3 &  9.2 &                   $0.26 \pm 0.14$ \\
 ASKAP~J005618.1$-$273012 & $14.07556$ & $-27.5035$ &  $2.006 \pm 0.559$ &  $1.770 \pm 0.066$ &  $2.613 \pm 0.070$ &  $2.050 \pm 0.069$ &  0.20 &    39.4 & 11.1 &                   $0.18 \pm 0.09$ \\
 ASKAP~J005709.0$-$243659 & $14.28753$ & $-24.6165$ &           $< 0.78$ &  $0.890 \pm 0.054$ &  $0.611 \pm 0.060$ &  $0.489 \pm 0.059$ &  0.31 &    13.5 & 14.2 &                   $0.22 \pm 0.10$ \\
 ASKAP~J005709.7$-$250751 & $14.29030$ & $-25.1310$ &           $< 0.81$ &  $0.654 \pm 0.054$ &  $0.814 \pm 0.062$ &  $0.447 \pm 0.062$ &  0.29 &    8.85 &  -- &                                -- \\
 ASKAP~J005729.6$-$231608 & $14.37350$ & $-23.2690$ &           $< 0.98$ &  $0.620 \pm 0.060$ &  $0.803 \pm 0.065$ &  $0.495 \pm 0.064$ &  0.24 &    5.76 &  -- &                                -- \\
 ASKAP~J005809.0$-$273407 & $14.53757$ & $-27.5688$ &           $< 0.79$ &  $0.849 \pm 0.068$ &  $0.602 \pm 0.072$ &  $0.552 \pm 0.073$ &  0.24 &    5.25 &  -- &                                -- \\
 ASKAP~J010004.6$-$231155 & $15.01934$ & $-23.1988$ &           $< 0.79$ &  $1.002 \pm 0.067$ &  $0.767 \pm 0.073$ &  $0.642 \pm 0.070$ &  0.23 &    7.15 &  -- &                                -- \\
 ASKAP~J010258.6$-$265119 & $15.74436$ & $-26.8555$ &           $< 0.87$ &          $< 0.099$ &  $0.261 \pm 0.091$ &  $0.232 \pm 0.098$ &  0.45 &    3.75 &  -- &                                -- \\
 ASKAP~J010534.6$-$231604 & $16.39415$ & $-23.2680$ &           $< 0.85$ &          $< 0.087$ &  $0.485 \pm 0.140$ &  $0.718 \pm 0.146$ &  0.58 &    3.36 &  -- &                                -- \\
\enddata
\tablenotetext{a}{There are 3 optical sources within 20\arcsec of this candidate. The two closest have a photometric redshift that is inconsistent with the distance to S190814bv.}
\tablenotetext{b}{Spectroscopic redshift.}
\end{deluxetable*}

\section{Untargeted Search for Radio Transients and Variables}\label{s_transients}

To search for a radio counterpart to S190814bv, we performed an untargeted search for transients and highly variable sources using the LOFAR Transients Pipeline \citep[TraP;][]{2015A&C....11...25S}.
We ran TraP with source detection and analysis thresholds of $5\sigma$ and $3\sigma$ respectively and used the `force beam' option to constrain the Gaussian shape fit parameters for all sources to be the same as the restoring beam. 

We selected candidates by identifying sources that were significant outliers in both variability metrics calculated by TraP: $\eta$, which is the weighted reduced $\chi^{2}$, and the variability index $V$ (equivalent to the fractional variability).
This was done by fitting a Gaussian function to the distributions of both metrics in logarithmic space, with $\sigma$ thresholds chosen to be $\eta > 1.5\sigma_\eta$ and $V > 1.0\sigma_{V}$, equating to values of $\eta > 2.73$ and $V > 0.18$. The thresholds were adapted from \citet{2019A&C....27..111R}, which gives approximate recall and precision rates of 90\% and 50\% respectively. 

This resulted in 285 transient or variable candidates, which was reduced to 89 sources after manual inspection to remove imaging artefacts and components of complex extended sources.

\subsection{Analysis of candidates for possible association with S190814bv}
The 89 variable sources were filtered to remove those that were not consistent with the predicted emission of S190814bv, which should not exhibit more than a single rise and decline on these timescales \citep{2016ApJ...831..190H}, according to the following criteria:
\begin{enumerate}
    \item Sources that showed a decline between epochs~1 and 2, followed by a rise between epochs~2 and 3. 41 sources were excluded.
    \item Sources detected in RACS epoch~0 where epochs~1 and 2 had lower integrated flux values than epoch~0. 3 sources were excluded.
\end{enumerate}

We then searched the GLADE catalogue \citep[GLADE;][]{2018MNRAS.479.2374Z} for galaxies in the localisation volume within 20\arcsec \citep[or $\sim 20$\,kpc at the estimated distance of S190814bv][]{GCN25333} of a variable source. We found one candidate (ASKAP~J005547.4$-$270433) that is near \object[2dFGRS TGS211Z177]{2dFGRS TGS211Z177}, a catalogued galaxy with $z=0.0738$ \citep{2001MNRAS.328.1039C}. This source was the only strong candidate after epoch 2 and prior to the acquisition of epoch 3 we performed multi-wavelength follow-up which we discuss in Section \ref{s_follow}. We excluded two candidates that matched with a GLADE galaxy $>3\sigma$ beyond the estimated distance to S190814bv \citep[$267 \pm 52$\,Mpc][]{GCN25333}.

We crossmatched the 42 remaining variable candidates with the Photometric Redshifts for the Legacy Surveys (PRLS) catalogue (Zhou et al. in prep.), which is based on Data Release 8 of DESI Legacy Imaging Surveys \citep{2019AJ....157..168D}. We excluded 22 variable sources that had all optical matches at distances differing by $>3\sigma$ from the estimated distance to S190814bv. This left 7 sources with at least one crossmatch within the localisation volume and 13 sources with no reliable distance estimate (see Table \ref{tab:candidates}).

\begin{deluxetable}{lccc}
\tablecaption{Radio observations of \object[AT2019osy]{AT2019osy}. Observations with the ATCA and VLA were carried out with maximum baselines of 6\,km and 40\,km respectively.\label{tab:at2019osy_radio}}
\tablecolumns{4}
\tablehead{
\colhead{Telescope} &
\colhead{$\Delta $T} &
\colhead{Frequency} &
\colhead{Flux Density}\\
\colhead{} &
\colhead{(days)} &
\colhead{(GHz)} &
\colhead{($\mu$Jy)}\\
}
\startdata
        ASKAP & 2 & 0.943 & 376 $\pm$ 33 \\
        \hline
        ASKAP & 9 & 0.943 & 550 $\pm$ 34 \\
        \hline
        VLA & 13 & 1.5 & 409 $\pm$ 34 \\
         & & 3.0 & 301 $\pm$ 21\\
         & & 6.0 & 213 $\pm$ 11\\
         & & 10.0 & 187 $\pm$ 11\\
        \hline
        ATCA & 14 & 5.0 & 369 $\pm$ 23 \\
         & & 6.0 & 335 $\pm$ 19\\
         & & 8.5 & 307 $\pm$ 15\\
         & & 9.5 & 278 $\pm$ 14\\
        \hline
        ATCA & 22 & 5.0 & 380 $\pm$ 21 \\
         & & 6.0 & 353 $\pm$ 17\\
         & & 8.5 & 299 $\pm$ 14\\
         & & 9.5 & 234 $\pm$ 14\\
        \hline
        VLA & 25 & 1.5 & 303 $\pm$ 48\\
         & & 3.0 & 317 $\pm$ 21\\
         & & 6.0 & 220 $\pm$ 10\\
         & & 10.0 & 150 $\pm$ 10\\
        \hline
        ASKAP & 33 & 0.943 & 513 $\pm$ 34\\
        \hline
        ATCA & 34 & 5.0 & 348 $\pm$ 17\\
         & & 6.0 & 349 $\pm$ 14\\
         & & 8.5 & 320 $\pm$ 15\\
         & & 9.5 & 275 $\pm$ 14\\
     \enddata
\end{deluxetable}

\section{Follow-up of ASKAP J005547.4$-$270433}\label{s_follow}
\subsection{Radio Observations}
We carried out follow-up observations of \object[ASKAP~J005547.4$-$270433]{ASKAP~J005547.4$-$270433} (hereafter \object[AT2019osy]{AT2019osy}) with the ATCA (C3278, PI: Dobie) using two 2\,GHz bands centered on 5.5 and 9\,GHz at 14, 22 and 34 days post-merger. We reduced the data using the same method as \citet{2018ApJ...858L..15D} using \object[PKS B1934-638]{PKS~B1934$-$638} and \object[B0118$-$272]{B0118$-$272} as flux and phase calibrators respectively. 

We also carried out VLA observations (VLA 18B-320, PI: Frail) on 2019 Aug 28 and Sep 09. Standard 2 bit WIDAR correlator setups were used for L and S bands, and 3 bit setups for C and X bands to obtain a contiguous frequency coverage between $1-12$~GHz. \object[3C48]{3C48} and \object[J0118$-$2141]{J0118$-$2141} were used as the flux and phase calibrators respectively. The data were processed using the NRAO CASA pipeline and imaged using the \texttt{clean} task in CASA.

A summary of our observations is given in Table~\ref{tab:at2019osy_radio}. We find a flux density offset\footnote{The flux densities of nearby sources and the calibrator source J0118-2141 between the ATCA and the VLA are consistent with the flux offset of ~40\% seen in AT2019osy. This offset can partially be explained by resolution effects, and detailed investigation of it is ongoing.} of $\sim 40\%$ between the initial ATCA and VLA observations, however later observations with both telescopes are self-consistent. We therefore find no evidence for radio variability beyond the initial rise observed with ASKAP.

\subsection{Optical Observations}
We conducted optical imaging of \object[AT2019osy]{AT2019osy} with the Dark Energy Camera \citep[DECam,][]{Flaugher2015} on the 4m Blanco telescope under NOAO program ID 2019B-0372 (PI: Soares-Santos). Images including the location of \object[AT2019osy]{AT2019osy} were taken in $i$ and $z$ bands nightly from 2019-08-15 to 2019-08-18 and on 2019-08-21 (UT) and reduced in real-time \citep{GoldsteinAndreoni2019}. A detailed offline analysis of the subtraction images zooming in on the location around AT2019osy, reveals no robust point source at this location to a depth of $i>21.2$mag and $z>20.0$mag on UT 2019-08-15 (the night of the merger) increasing linearly in limiting magnitude to  $i > 23.5$mag and $z>23.5$mag on UT 2019-08-21 (consistent with independent analysis by \citealt{gcn25495}). We also analyzed the DECam images using \texttt{The Tractor} image modeling software \citep{Lang2016} and found that a model with an exponential galaxy profile with a point source at the galaxy nucleus is required to fit the data, both before and after S190814bv. This suggests that there is no optical transient temporally coincident with S190814bv but possibly some underlying nuclear variability.    

On 2019-08-22 UT, we observed \object[AT2019osy]{AT2019osy} in the near infrared using the Wide-field Infrared Camera \citep[WIRC,][]{Wilson2003} with the 200-inch Hale telescope at Palomar Observatory for a total of 10 minutes exposure time \citep{gcn25449}. The WIRC data were reduced and stacked using a custom pipeline (De et al., in preparation). No counterpart to \object[AT2019osy]{AT2019osy} was detected down to an AB limiting magnitude of $J > 21.5$ ($5 \sigma$).

We also obtained a spectrum of the host galaxy of \object[AT2019osy]{AT2019osy} using the Double Beam Spectrograph \citep{OkeGunn1982} on the Palomar 200-inch Hale Telescope (P200), which we reduced using \texttt{pyraf-dbsp} \citep{pyrafdbsp}. The spectrum is dominated by red continuum that is likely primarily associated with the host galaxy; no obvious broad features are evident. We identify several narrow emission lines (H$\alpha$; [NII]$\lambda \lambda$6548,6583, [SII]$\lambda \lambda$6716,6731, and marginal [OII]$\lambda$3727) at a common redshift of 0.0733, consistent within 2-sigma of the LVC distance constraint.  H$\beta$ and [OIII]$\lambda$5007 are not detected in the spectrum. We measure a flux ratio of log[NII$\lambda$6583/H$\alpha$]=0.2, indicating at least partial contribution by an AGN \citep{Kauffmann+2003}.

\subsection{X-ray observations}
We observed the field of \object[AT2019osy]{AT2019osy}, starting at 2019-09-23 10:30:48 UT for 20\,ks with the \texttt{Chandra} ACIS-S instrument (S3 chip) and very faint data mode. The data were analyzed with \texttt{CIAO} \citep[v 4.11;][]{2006SPIE.6270E..1VF} and calibration was carried out with \texttt{CALDBv4.8.4.1}. We reprocessed the primary and secondary data using the \texttt{repro} script, created X-ray images for the 0.3--8\,keV  range.  No sources were visible near \object[AT2019osy]{AT2019osy} (verified with both \texttt{wavdetect} and \texttt{celldetect}), with a maximum count rate of 2.85$\times10^{-4}\,{\rm s}^{-1}$. Assuming a neutral hydrogen column density $N_{\rm H}=1.8\times10^{20}\,{\rm cm}^{-2}$ and a power-law model with index $n=1.66$ (corresponding to the observed radio spectral index of $-0.4$), this count rate yields a 0.3--8\,keV unabsorbed flux upper limit of 3.2$\times10^{-15}\,{\rm erg\,cm}^{-2}\,{\rm s}^{-1}$ \citep[as reported in][]{GCN25822} or an unabsorbed luminosity of 4.2$\times10^{40}\,{\rm erg\,s}^{-1}$.

\subsection{Source classification}
\object[AT2019osy]{AT2019osy} exhibits no significant radio variability beyond the initial rise and there is no evidence for a coincident optical transient. The coincident galaxy is edge-on, likely with significant dust obscuration towards the nucleus, and therefore the optical spectrum is consistent with an AGN within a star-forming galaxy. The inferred radio and X-ray luminosity of \object[AT2019osy]{AT2019osy} along with the small offset from the optical centroid of \object[2dFGRS TGS211Z177]{2dFGRS TGS211Z177} suggests that the source is a variable low-luminosity AGN \citep{2012A&A...545A..66B} and unrelated to S190814bv.

\section{Discussion}

\subsection{Candidate classification}
\label{sec:candidate_classification}
We find 21 candidate counterparts to S190814bv above a threshold of $170\,\mu$Jy, which is consistent with the expected rate of AGN variability \citep{2019arXiv190912588R}. Additionally, the expected level of compact source variability caused by refractive interstellar scintillation along this line of sight is $\sim 35\%$ \citep{2002astro.ph..7156C}, comparable to $V_{\rm int}$ for all but three sources which we discuss below.

We classify \object[ASKAP J005434.6$-$280235]{ASKAP J005434.6$-$280235} as a variable AGN based on follow-up observations \citep{gcn25449,GCN25445}. \object[ASKAP J010258.6$-$265119]{ASKAP J010258.6$-$265119} is coincident centrally between two large radio lobes and hence likely associated with core emission from a radio galaxy. \object[ASKAP J010534.6$-$231604]{ASKAP J010534.6$-$231604} is coincident ($<1$\arcsec) with \object[WISE J010534.64$-$231605.5]{WISE J010534.64$-$231605.5} \citep{2012yCat.2311....0C}, which is likely a variable AGN at a distance of $z\sim 1$ \citep{2017arXiv170908634G}.

While we cannot conclusively rule the sources in Table \ref{tab:candidates} out as counterparts to S190814bv, they are likely AGN exhibiting a combination of intrinsic and extrinsic variability. Of course, at most one candidate can be the actual counterpart, and there is nothing yet to distinguish any of these from the others. Further observations on timescales of months--years will reveal their nature.

\subsection{Radio transient rates}
Our follow-up of S190814bv is the most sensitive widefield radio transients search to-date, approximately an order of magnitude more sensitive compared to previous searches with comparable areal coverage \citep{2016MNRAS.456.3948H} and approximately an order of magnitude more areal coverage than previous searches at comparable sensitivities \citep{2013ApJ...768..165M}.

We have found 4 transient candidates (i.e. sources with a prior constraining non-detection) in total; the three sources discussed in Section \ref{sec:candidate_classification} and \object[ASKAP J005104.2$-$230852]{ASKAP J005104.2$-$230852}, which was ruled out as a candidate to S190814bv based on the redshift of nearby optical sources. This source is coincident ($<0.6$\arcsec) with WISE J005104.13-230851.8, which  is likely a variable AGN.

We therefore place an upper-limit on the 943\,MHz radio transients surface density of $0.05\,\deg^{-2}$ for sources above $170\,\mu{\rm Jy}$ at 95\% confidence.

\subsection{Non-detection of a radio afterglow from S190814bv}
Predicted radio lightcurves from NSBH mergers span a large range of flux densities and timescales \citep[e.g.][]{2013MNRAS.430.2121P,2016ApJ...829..112L,2019MNRAS.486.5289B}. If the radio emission is dominated by the outflowing dynamical ejecta the lightcurve will peak on timescales of years, whereas if the emission is jet-dominated the lightcurve will peak at comparably lower flux densities on timescales of days--months \citep{2016ApJ...831..190H}. In each of these scenarios the lightcurve is also dependent on the merger energetics, circum-merger density and inclination angle, each of which can change both the peak time and flux density by an order of magnitude. The merger energetics are determined by the mass ratio, the spin of the black hole (both of which are calculable from gravitational wave strain data that is yet to be released) and the unknown neutron star equation of state \citep{2011PhRvD..84f4018K,2012PhRvD..86l4007F}.

\begin{figure}
    \centering
    \includegraphics[width=\linewidth]{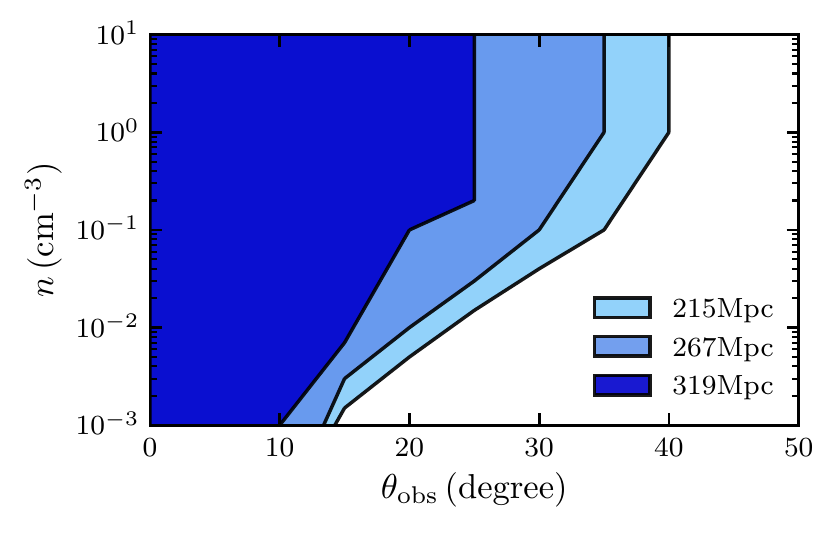}
    \caption{Radio constraints on viewing angle and circum-merger density for a merger with isotropic equivalent energy $10^{51}$\,erg and an initial jet opening angle of $10\degr$. Shaded regions correspond to parts of the parameter space that are ruled out by our radio constraints for a range of distances corresponding to $1\sigma$ either side of the median.}
    \label{fig:merger_constraints}
\end{figure}

We place a $5\sigma$ upper limit on the 943\,MHz radio emission from S190814bv of $170\,\mu$Jy at $\Delta T= 2$, 9 and 33 days post-merger. Figure \ref{fig:merger_constraints} shows the constraints we can place on the merger inclination angle, $\theta_{\rm obs}$, and circum-merger density, $n$, assuming the afterglow has an isotropic equivalent energy $E_{\rm iso}=10^{51}$\,erg \citep[typical of short GRB afterglows;][]{2015ApJ...815..102F}, an initial jet opening angle of $10\degr$ and microphysics parameters $\epsilon_{e}=0.1$, $\epsilon_{B}=0.01$ and $p=2.2$. We can rule out the part of the parameter space typically occupied by short GRBs, assuming that their inclination angle is smaller than the opening of the angle of the jet \citep{2015ApJ...815..102F}. Under a more conservative assumption of the isotropic equivalent energy ($E_{\rm iso}=10^{50}$\,erg) we can only rule out a small part of the parameter space around $\theta_{\rm obs}=10\degr$ and $n=1\,$cm$^{-3}$.

In comparison, if we scale the non-thermal lightcurve of GW170817 to 943\,MHz based on a spectral index of $\alpha=-0.575$ \citep{2018ApJ...868L..11M,2019arXiv190906393H} and place it at a distance comparable to S190814bv, we find a peak flux density of $\sim 5\,\mu$Jy, well below our detection threshold. We note that the non-thermal emission  from GW170817 did not peak until $\sim 150\,$d post-merger \citep{2018ApJ...858L..15D}. Further observations on timescales of months--years post-merger will enable us to place tighter constraints on the circum-merger density and inclination angle, which may be useful in improving the gravitational wave localisation \citep{2019MNRAS.488.4459C}.

\section{Conclusions}
We have performed widefield radio follow-up of the NS-BH merger S190814bv with the Australian Square Kilometre Array Pathfinder. We cover 89\% of the sky localisation with a single $30\,\deg^2$ pointing centered on the localisation maxima. We found 21 candidate counterparts and performed comprehensive multi-wavelength follow-up of one, AT2019osy. The number of candidates is consistent with the expected rate of AGN variability. Most exhibit variability that is consistent with that expected from interstellar scintillation and are therefore unlikely to be related to S190814bv

The non-detection of a radio counterpart allows us to place constraints on the circum-merger density, $n$, and inclination angle of the merger, $\theta_{\rm obs}$. Under the assumption of $E_{\rm iso} = 10^{51}$\,erg, we constrain $\theta_{\rm obs} > 10\degr$ for all $n$ at the extreme of the probability distribution of distance to the event. We will be able to place tighter constraints on these merger parameters once inclination angle estimates from gravitational wave strain data are released publicly.

As well as probing different parameters to optical searches, radio observations of future events may detect a gravitational wave counterpart where optical follow-up is inhibited by observing constraints, or intrinsic properties of the merger. We have demonstrated that it is possible to perform comprehensive follow-up of gravitational wave events with ASKAP, due to its large field of view that enables a survey speed significantly faster than comparable radio facilities.

\section*{Acknowledgements}
DD is supported by an Australian Government Research Training Program Scholarship. TM acknowledges the support of the Australian Research Council through grant DP190100561. DLK and IB were supported by NSF grant AST-1816492. Parts of this research were conducted by the Australian Research Council Centre of Excellence for Gravitational Wave Discovery (OzGrav), project number CE170100004. We acknowledge support by the GROWTH (Global Relay of Observatories Watching Transients Happen) project funded by the National Science Foundation PIRE (Partnership in International Research and Education) program under Grant No 1545949.A.C. acknowledges support from the NSF CAREER award \#1455090. D. A. Goldstein acknowledges support from Hubble Fellowship grant HST-HF2-51408.001-A. Support for Program number HST-HF2-51408.001-A is provided by NASA through a grant from the Space Telescope Science Institute, which is operated by the Association of Universities for Research in Astronomy, Incorporated, under NASA contract NAS5-26555. Development of the PRLS photometric redshift catalog used here was supported by the U.S. Department of Energy, Office of Science, Office of High Energy Physics under award number DE-SC0007914. The National Radio Astronomy Observatory is a facility of the National Science Foundation operated under cooperative agreement by Associated Universities, Inc.

The Australian SKA Pathfinder is part of the Australia Telescope National Facility which is managed by CSIRO. Operation of ASKAP is funded by the Australian Government with support from the National Collaborative Research Infrastructure Strategy. ASKAP uses the resources of the Pawsey Supercomputing Centre. Establishment of ASKAP, the Murchison Radio-astronomy Observatory and the Pawsey Supercomputing Centre are initiatives of the Australian Government, with support from the Government of Western Australia and the Science and Industry Endowment Fund. We acknowledge the Wajarri Yamatji people as the traditional owners of the Observatory site.

The Australia Telescope Compact Array is part of the Australia Telescope National Facility which is funded by the Australian Government for operation as a National Facility managed by CSIRO.

This research has made use of NASA's Astrophysics Data System Bibliographic Services.

\facility{ASKAP, ATCA, VLA, DECam, P200, WIRC, Chandra}
\software{
ASKAPsoft \citep{2017ASPC..512..431W},
BANE \citep{2018PASA...35...11H},
CASA \citep{2007ASPC..376..127M}, 
pyraf-dbsp \citep{pyrafdbsp}, 
The Tractor \citep{Lang2016}, 
TraP \citep{2015A&C....11...25S}}.


\end{document}